# Insights into the Mechanism underlying the Chiral-Induced Spin Selectivity:
# The effect of an Angle-Dependent Magnetic Field and Temperature

Tapan Kumar Das, Ron Naaman

Department of Chemical and Biological Physics,

Weizmann Institute of Science, Rehovot 7610001, Israel

Jonas Fransson

Department of Physics and Astronomy, Uppsala University, Uppsala 75236, Sweden

**Abstract**

Chiral oligopeptide monolayers were adsorbed on a ferromagnetic surface and their magnetoresistance was measured as a function of the angle between the magnetization of the ferromagnet and the surface normal. These measurements were conducted as a function of temperature for both enantiomers. The angle dependence was found to follow the cosine square function. Quantum simulations revealed that the angular distribution could be obtained only if the monolayer has significant effective spin orbit coupling (SOC), that includes contribution from the vibrations. The model shows that SOC only in the leads cannot reproduce the observed angular dependence. The simulation can reproduce the experiments only if it included electron-phonon interactions and dissipation.

The chiral-induced spin selectivity (CISS) effect is by now a well-established phenomenon.[1] However, theoreticians still debate about the mechanism underlying this effect.[2] Part of the confusion results from the notion that spin-orbit coupling (SOC) in chiral hydrocarbon-based molecules is extremely small. Nevertheless, very efficient spin polarization has been observed in those systems.[3,4,5] Since a significant magnitude of an effective SOC must be an essential property for the large effect to exist, at room temperature,[6] in several theoretical models different ways to introduce it were proposed.[7,8,9] Quite early, after the effect was observed, it has been suggested that the substrate on which the chiral molecule is adsorbed is the source of the SOC and that actually the chiral molecule serves as an orbital angular momentum filter.[10,11] Several experiments were conducted to probe the role of the substrate's SOC by varying the substrates.[12] However, the results obtained were not conclusive enough to determine the role of the SOC, since the substrate may have other effects, like the location of the Fermi level relative to the gap between the highest occupied and the lowest unoccupied molecular orbitals and polarizability, which can determine the conductivity and the barrier at the interface and therefore influence the observed spin polarization.

In spintronics, in which spins are injected from ferromagnetic substrates, studies were performed that probed the dependence of the spin polarization on the angle θ between the magnetization of the ferromagnet and the electric field used to drive the current. The angle dependence results from the anisotropy of the magneto resistance.[13] Typically, it was found that the spin polarization depends on $\cos^2\theta$.[14,15]

In the present work, we investigated the dependence of the magnetoresistance on the magnet's angle at various temperatures. The spin-valve device used for measuring the magnetoresistance (MR) consisted of four contacts and was fabricated in a crossbar geometry (**Figure 1a**). The bottom electrode was made from titanium (Ti), followed by gold (Au) on which the chiral molecules were adsorbed. The molecules are D-Al5 and L-Al5, where Al5 stands for SHCH$_2$CH$_2$CO-{Ala-AiB}$_5$-COOH, Ala refers to alanine, and AiB refers to 2-aminoisobutyric acid. Following the adsorption, 1.5 nm of MgO was deposited, followed by a ferromagnetic nickel (Ni) and a thin layer of gold. A magnetic field (B) up to ±1 T was applied to measure the magnetoresistance dependence on θ (**Figure 1b**) and its temperature dependence. The resistance of oligopeptides was measured by applying 1 mA current perpendicular to the sample surface, while the magnetic field was rotated with respect to the direction of the input current. A schematic of the mechanism for spin transport through helical molecules is shown in **Figure S1a**. There is an electric field pointing perpendicular to the velocity of the electron that confines the electron that moves within the helical potential. The electric field acts on the electron on the left or right, relative to the velocity, depending on the handedness of the molecule. The MR is plotted as a function of magnetic field for the D-Al5 and L-Al5 molecules at different temperatures shown in **Figures S1b** and **S1c**, respectively. The MR is defined as $MR\ (\%) = \frac{R(B)-R(0)}{R(0)} \times 100$, where R(B) and R(0) are the resistances measured at a magnetic field $B$ and with no magnetic field, respectively. The spectra were collected by scanning the magnetic field between -1.0 and 1.0 T at different angles and temperatures.

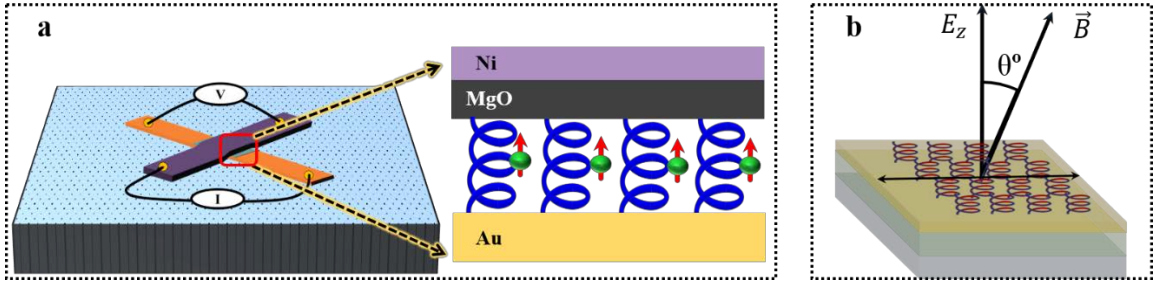

**Figure 1:** a) Schematic of a spin-valve device for measuring magnetoresistance. The interface of the device is magnified and shown in the magnified image of (a), which explains how oligopeptide is sandwiched between the bottom and top electrodes. b) A schematic of an angle-dependent magnetoresistance (MR) measurement, where the magnetic field B is applied at different angles with respect to the current, $E_z$, perpendicular to the molecular surface.

**Figure 2** presents the angle-dependent magnetoresistance curve, when a field of 0.5 T is applied at different temperatures. For D-Al5, the angle-dependent magnetoresistance, R, was fitted (**Figure 2c**) by the equation

$$R = R_\perp + (R_\parallel - R_\perp)\sin^2\theta/2, \quad (1)$$

where $R_\parallel\ and\ R_\perp$ are the resistance, when the angle between the magnetization and the direction of current is 0° or 90°, respectively. The magnetoresistance shows a changing trend with a period of 360°. For L-Al5 the angle-dependent magnetoresistance was fitted (**Figure 2d**) with the equation,

$$R = R_\perp + (R_\parallel - R_\perp)\cos^2\theta/2 \quad (2)$$

An interesting observation is the increase in the MR values with temperature. This temperature effect was observed before and was explained by the contribution of the low-frequency vibrations to the effective SOC.[16] Although the values of the MR increase with temperature, the angle dependence remains identical at all temperatures. All the fits are extremely good.

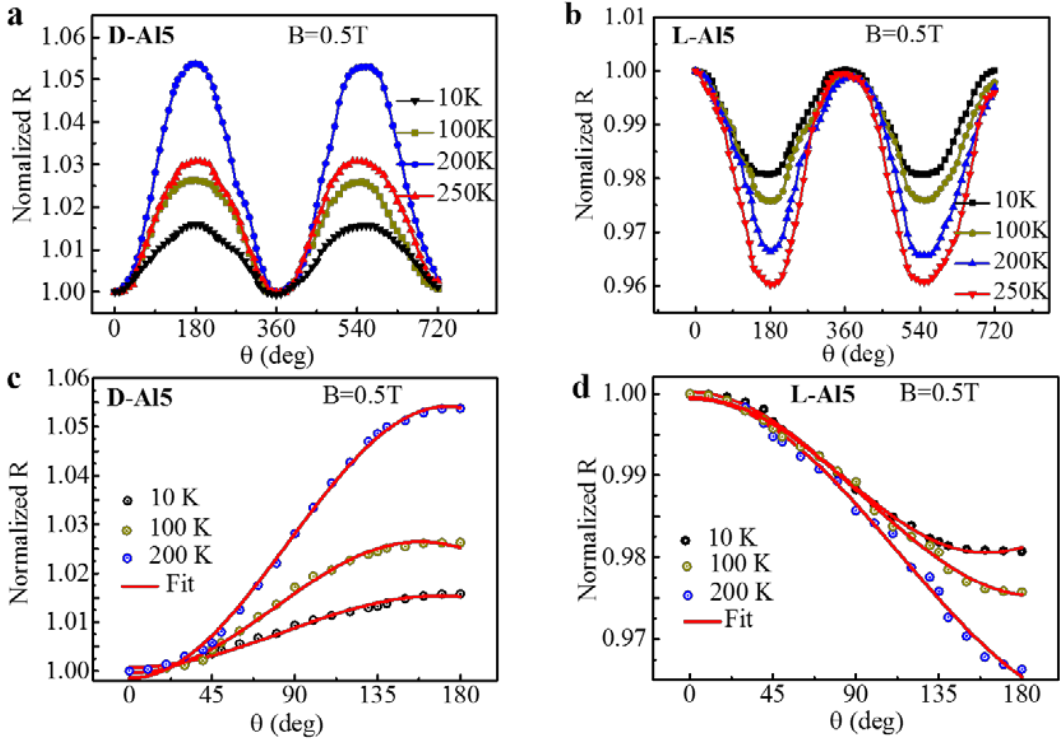

**Figure 2:** The angle-dependent resistance measured at four different temperatures with a 0.5T applied magnetic field for (a) D-Al5 and (b) L-Al5. The resistances are normalized at θ=0°. The resistance curves were fitted with two fitting curves from an angle between θ=0° and θ=180°; the fitting curve $R = R_\perp + (R_\parallel - R_\perp)\cos^2 \theta/2$ was used for L-Al5 and $R = R_\perp + (R_\parallel - R_\perp)\sin^2 \theta/2$ for D-Al5, respectively. For all the fittings, the fit parameters are $R^2$=0.99 and $\chi^2$=0.001 to 0.0005.

It is important to understand that in the case of the CISS effect, the MR curves are asymmetric, unlike the common MR effect in spintronics. This effect is due to the transport medium, namely, the chiral molecules, which have different resistances for the same spin moving in two opposite directions. **Figure 3** summarizes all the measurements performed on the two enantiomers. It presents the magnetic field-dependent resistance at different temperatures (**Figure 3a,b**). **Figures 3c,d** show the resistance when θ equals 0° and 180°, indicating the relative directions of the current and the magnetic field, respectively. Figures 3e,f show the dependence of the resistance on θ for three different temperatures. Detailed studies of the angle-dependent MR for both enantiomers are shown in **Figure S2-S7** for different temperatures.

The magnitudes of the MR are relatively low, compared with measurements performed using magnetic contact AFM.[17] This is due to the relatively large area of the present devices, which results in pinholes in the self-assembled monolayers. The current leaking through these pin holes faces less resistance and is not spin selective. Hence, it creates a background of unpolarized spin current that reduces the magnetoresistance.

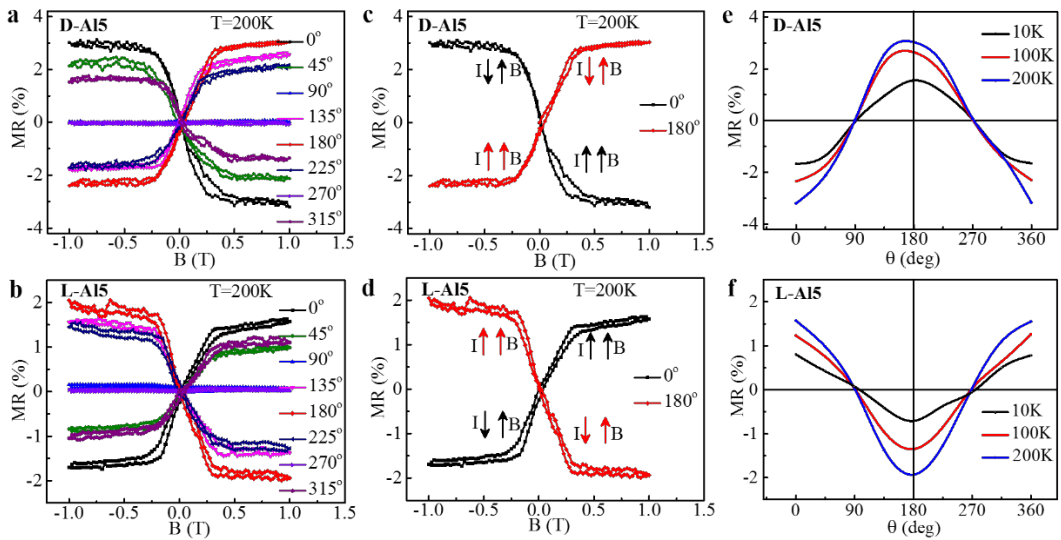

**Figure 3:** The angle-dependent MR measured at T=200K as a function of a magnetic field scanning from -1T to +1T for (a) D-Al5 and (b) L-Al5. The comparative MR graphs measured for angles θ=0º and θ=180º at 200K show how MR changes in (c) for D-Al5 and (d) for L-Al5. The angle-dependent MR is calculated for a magnetic field of 0.5T for three different temperatures and is plotted in (e) D-Al5 and (f) L-Al5. Current of I=1mA was applied in all the measurements and the magnetic field, B, was applied at different angles when the electric field, $E_z$, was applied perpendicular to the surface on which the molecules are adsorbed.

For determining the relationship between the experimental observations and the CISS mechanism, we present the results from simulations using several different models to compare and analyze different perspectives, which may serve as a viable source for the chiral-induced spin selectivity effect. Therefore, we used the multi-orbital model introduced in Ref. [18] and the model with vibrationally enhanced spin-orbit coupling presented in Ref. [19]. We will refer to the former as Model 1 and the latter as Model 2 (see **Figure 4**a). Both models comprise $\mathcal{M} = M \times N$ sites distributed on the coordinates $\boldsymbol{r}_m$, forming a helix with $M$ laps and $N$ sites per lap, where each site $m$ is represented by $n$ electron energy levels $\epsilon_{m\nu}$, $\nu = 1, 2, \ldots, n$. These levels are mixed by hybridization, represented by the parameter $\boldsymbol{w}_{m\nu\nu'} = w_0 \sigma^0 + \boldsymbol{w}_1 \cdot \boldsymbol{\sigma}$, where $w_0$ and $\boldsymbol{w}_1$ denote the spin-conservative and spin-mixing components, respectively, of this hybridization, and $\boldsymbol{\sigma}$ is the vector of Pauli matrices. Each level $\nu$ on site $m$ is coupled to the level $\nu'$ on its nearest-neighboring sites, $m \pm 1$, via the inter-site hybridization $\boldsymbol{t}_{m\nu\nu'} = t_0 \sigma^0 + \boldsymbol{t}_1 \cdot \boldsymbol{\sigma}$, where $t_0$ and $\boldsymbol{t}_1$ denote the spin-conservative and spin-mixing components, respectively. The chirality of the structure is captured by the vectors $\boldsymbol{v}_m^{(+)} = \hat{\boldsymbol{d}}_m \times \hat{\boldsymbol{d}}_{m+1}$ and $\boldsymbol{v}_m^{(-)} = -\boldsymbol{v}_{m-2}^{(+)}$, where $\boldsymbol{d}_m = \boldsymbol{r}_m - \boldsymbol{r}_{m+1}$, $\hat{\boldsymbol{d}}_m = \boldsymbol{d}_m/|\boldsymbol{d}_m|$; they measure the curvature between the next-nearest-neighboring sites. In Model 1 the chirality of the molecule is associated with an effective spin-orbit coupling, $\lambda_1 \boldsymbol{v}_m^{(+)} \cdot \boldsymbol{\sigma}$, and orbital coupling, $\lambda_0 v_{mz}^{(+)} \sigma^0$; the latter is introduced to also include chirality in the absence of structurally generated spin-orbit coupling. It is worth mentioning that in the simulations we made with the spin-orbit coupling coupled to the chirality, the parameter $\lambda_1$ is non-zero and $\lambda_0$ is set to zero. For simulations where chirality is not connected to the spin-orbit coupling, we set $\lambda_1 = 0$ and $\lambda_0 \neq 0$. In Model 2 a vibrational component, with vibrational energy $\omega_{vib}$, is connected to the nearest neighbor and next-nearest-neighbor hopping with rates $t_{vib}$ and $\lambda_{vib}$, respectively, where the latter represents vibrationally induced spin-orbit coupling.

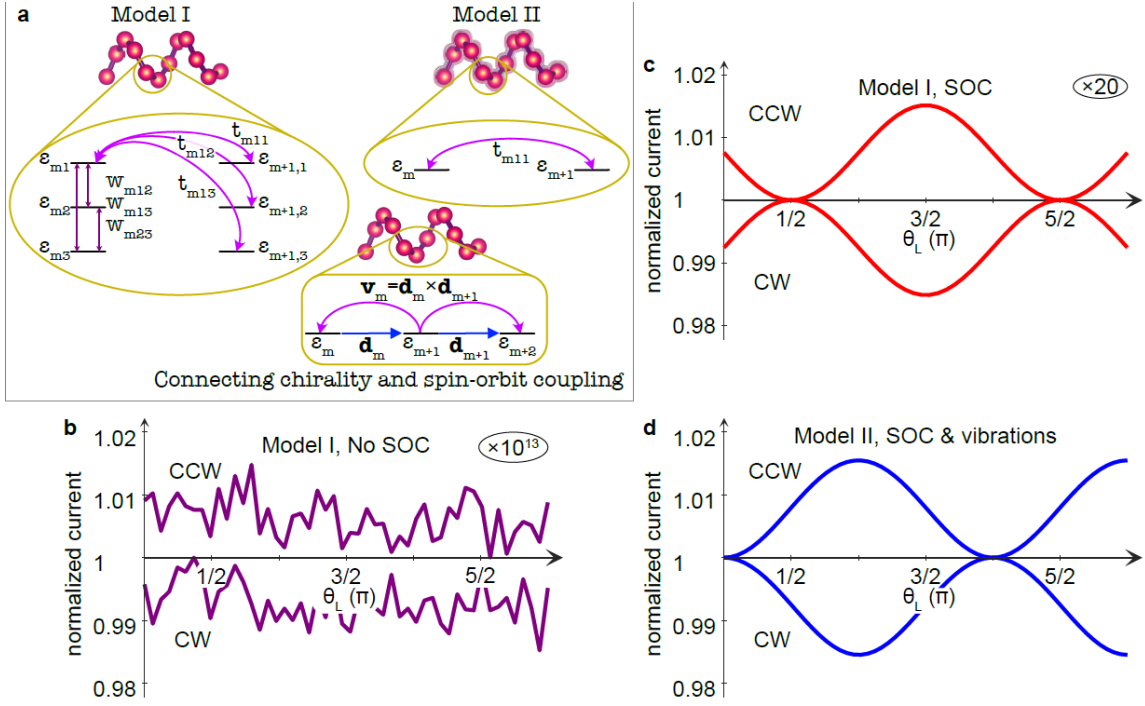

**Figure 4**: a) Schematics of the set-up in Model I (left) and Model II (right), respectively, and the next-nearest-neighbor hopping that couples to the chirality of the structure (bottom). b) – d) Angular dependence of the charge current for counter-clockwise (above one) and clockwise (below one) helices. The currents were calculated using Model 1 b) without ($\lambda_1 = 0$) and c) with ($\lambda_0 = 0$) spin-orbit coupling, and d) Model 2. In Model 1, we used the next-nearest hopping $t_0 = 1$ eV; the other parameters are $(\epsilon_{m1}, \epsilon_{m2}) = -(2, 0.5)t_0$, $w_0 = t_0/10$, $\lambda_0 = \lambda_1 = t_0 \cdot 10^{-3}$, $p_R = 0.2$, $\phi_L = \pi$, $\phi_R = \pi/2$, $\theta_R = \pi/2$. In Model 2, we used $\epsilon_m = -2t_0$, $\lambda_1 = t_0 \cdot 10^{-3}$, $t_{vib} = t_0/10$, $\lambda_{vib} = t_0 \cdot 10^{-4}$, $\omega_{vib} = t_0 \cdot 10^{-6}$, and $\tau_{vib} = 2/5t_0$. For all simulations we used $\Gamma_0 = t_0/10$, $p_L = 0.2$, $V_{sd} = 4t_0/e$, and $T = 300$ K. Note that the currents in b) and c) are multiplied by $10^{13}$ and 20 to be on the same scale as the current in d).

Using these models, the calculations of the charge current $J$ follow the procedure described in Refs. [18,19] with the molecule sandwiched between a non-magnetic and a ferromagnetic metallic lead, where the magnetic moment of the latter aligns with the angle $\theta$ relative to the direction of the current. In addition, the couplings $\boldsymbol{\Gamma}^\chi$ to the left ($\chi = L$) and right ($\chi = R$) leads are parametrized using $\boldsymbol{\Gamma}^\chi = \Gamma_0(\sigma^0 + \boldsymbol{p}_\chi \cdot \boldsymbol{\sigma})/2$, where $\boldsymbol{p}_\chi = p_\chi(\cos\phi_\chi \sin\theta_\chi, \sin\phi_\chi \sin\theta_\chi, \cos\theta_\chi)$, $p_\chi \in [-1,1]$, to account for possible spin-orbit couplings in the leads, which are transferred to the molecule through the interface.

It was demonstrated in Ref. [18] that the multi-orbital model can be used to reproduce the CISS effect albeit not with the correct magnitude. In fact, the magnitude of the normalized difference $100 \cdot (J_\uparrow - J_\downarrow)/(J_\uparrow + J_\downarrow)$ is less than 1 % for all choices of parameters, where the subscript $\uparrow$ ($\downarrow$) corresponds to the angle $\theta = 0$ ($\theta = \pi$) in the current context. In contrast, in non-equilibrium, the model has the correct symmetry properties along with changes in the spin polarization of the injected current. This is the property that we are interested in here; however, the correct magnitude of the effect must be addressed using other models. By adding, for instance, nuclear vibrations into the model, see Ref. 19, using Model II the amplitude is expected to become similar with the experimental results.

The plots in **Figure 4b-d** show the angular dependence of the charge current for angles $0 \leq \theta_L \leq 3\pi$, for the two enantiomers, clockwise (CW) and counter-clockwise (CCW). The plots in the three panels represent three different model configurations for each enantiomer. First, in **Figure 4b,** we used Model 1 and set the spin-orbit couplings $\boldsymbol{w}_1$, $\boldsymbol{t}_1$, and $\lambda_1$ to zero. Here, we assumed that there is a spin-orbit coupling in the leads, which leak into the molecule via the coupling $\boldsymbol{\Gamma}^\chi$, which is modeled by setting the angles $\phi_L = \pi$ and $(\phi_R, \theta_R) = (\pi/2, \pi/2)$, resulting in the appropriate anti-symmetry of the coupling matrices. However, despite a large spin-orbit coupling leaking through the interfaces between the metals and the molecule, there is no apparent effect on the charge currents, neither with respect to the enantioselectivity nor with respect to the angular dependence. Moreover, adding the local spin-orbit couplings $\boldsymbol{w}_1$ and $\boldsymbol{t}_1$ to the simulations (not shown) does not qualitatively change the result. The independence of the angular variations remains as well as the lack of enantiospecific signatures. It should be noticed that although we tried other angular configurations, the results remain independent of the rotation angle $\theta_L$. Thus, we can conclude that spin-orbit coupling in the leads by itself cannot result in the angular dependence observed and hence cannot be the source of the chiral-induced spin selectivity effect. This occurs despite the multi-orbital structure, which enables the development of orbital moments.

Second, in **Figure 4c** we extend Model I by adding the spin-orbit coupling $\lambda_1$; consequently, the currents become enantiospecific, as seen in the plots. The difference in this configuration compared to the previous one is that the spin-orbit coupling does not any longer consist of a simple local mixing of the spins but is instead coupled to the non-local extension of the structure. This means that even if the spins locally mix through the spin-orbit coupling, such mixing does not acquire any direction that pertains to the broken inversion symmetry of the structure. The directionality associated with the spin-orbit coupling may interplay with the directionality of the structure and either enhance or diminish the enantiospecific properties; however, this interplay can only be enabled if the spin-orbit coupling connects non-locally to the structure. Nevertheless, it can also be seen in **Figure 4c** that even if the currents are enantiospecific and show a clear response to the angular variations of the ferromagnetic lead, the angular variations are phase shifted by $\pi/2$ compared to the experimental results. This phase shift depends on the spin-orbit couplings in the leads as well as how they are phase shifted relative to each other. Despite this dependence, we have not been able to tune the spin-orbit couplings in the leads such that the phase shift of the currents vanishes. We therefore conclude that even with the spin-orbit coupling connected to the chirality, the elastic multi-orbital model does not capture the correct physics involved in the CISS effect.

The third simulation is based on Model 2, and as shown in **Figure 4a**, there is a single orbital per site, represented by the energy $\epsilon_m$, and where the sites are also connected by the vibrationally induced components $t_{vib}$ and $\lambda_{vib}$. Moreover, we have removed the spin-orbit coupling in the leads in order to focus on the effects provided by mechanism induced by the vibrations. The currents shown in **Figure 4d** are again enantiospecific, in agreement with previous results; they also vary with the angle of the magnetic moment in the ferromagnetic lead. In contrast to the results obtained with Model 1, the results from Model 2 do not have the erroneous phase shift and, as can be seen, they correctly reproduce the experimental results. We infer that this agreement arises due to the introduction of dissipation in the system connected to the lifetime of the electron levels that become accessible due to the vibrationally assisted couplings between the sites. Indeed, there is a lifetime, $\tau_{vib}$, associated with the vibrationally accessible electron levels, which in the simulation corresponds to a broadening energy of about 250 meV. Such a large broadening can be justified by the coupling between the vibrations and electrons as well as between the vibrations themselves, and physically represents how susceptible the system is to fluctuations. A large number of fluctuations suggests that the electrons can move around between different states, a motion that is constrained by the chiral structure and is, hence, enantiospecific.

We speculate that the dissipation associated with the vibrations gives rise to an enantiospecific anisotropy in the polarizability, which is sensitive to the spin of the injected electrons. The spin mixing caused by the spin-orbit interaction cannot by itself effectively give rise to strong changes in the transport properties upon changes in the spin-polarization of the injected electrons. Components of correlations and dissipation appear to be crucial for these types of anisotropic properties to emerge. Judging from this conclusion, we can also conjecture that the CISS effect is a spin filtering effect.

The results presented here lead to three important insights into the mechanism underlying the CISS effect. The results indicate that the observed angle-dependent magnetic field effect on the spin polarization cannot occur without the chiral system itself having SOC. The SOC in the lead by itself cannot create, according to the model presented here, the angular dependence. In addition, the effective SOC in the chiral system is not simply "atomic-like" SOC, but instead it pertains to the polarizability of the system and it results from contributions from electron-electron and electron-phonon interactions. Several polaron models suggested in the past are consistent with this conclusion.[20,21] Finally, the dissipation occurring through the interaction with phonons is essential for the observed angle dependence. This last observation is consistent with former theoretical suggestions.[22,23,24]

The establishment of these three effects is consistent with all the experimental observations related to the CISS. As suggested before, they show the importance of dissipation, the need to treat the effect in a treatment that goes beyond the single electron approximation by including electron-electron and electron-phonon interactions. The results, although not excluding the role of the surface-chiral system interface, indicate that the SOC in the lead, by itself, cannot result in the observed angle dependence. This is consistent with recent experimental results in which the CISS effect was observed in systems that are not attached to leads.

**Acknowledgments**

RN acknowledges partial support from a research grant from Jay and Sharon Levy, from the Sassoon and Marjorie Peress Philanthropic Fund, from the Estate of Hermine Miller, from the US Department of Energy Grant ER46430, and from the US-AFSOR Grant FA9550-21-1-0418.

**References**

[1] D.H. Waldeck, R. Naaman, Y. Paltiel, *APL Mater.* **2021**, 9, 040902.

[2] F. Evers, A. Aharony, N. Bar-Gill, O. Entin-Wohlman, P. Hedegård, O. Hod, P. Jelinek, G. Kamieniarz, M. Lemeshko, K. Michaeli, V. Mujica, R. Naaman, Y. Paltiel, S. Refaely-Abramson, O. Tal, J. Thijssen, M. Thoss, J. M. Van Ruitenbeek, L. Venkataraman, D. H. Waldeck, B. Yan, L. Kronik, Adv. Mater. **2022**, 34, 2106629.

[3] Q. Yang, Z. Zhang, X. Jiang, X. Wang, X. Wang, Z. Shang, F. Liu, J. Deng, T. Zhai, J. Hong, Y. Zhang, W. Zhao, IEEE Electron Device Lett. **2022**, 43, 1862.

[4] J. Labell, D. K. Bhowmick, A. Kumar, R. Naaman, T. Torres, Chem Sci. **2023**, 14, 4273.

[5] H. Al-Bustami, S. Khaldi, O. Shoseyov, S. Yochelis, K. Killi, I. Berg, E. Gross, Y. Paltiel, R. Yerushalmi, Nano Lett. **2022**, 22, 5022.

[6] P. Hedegård, J. Chem. Phys. **2023**, 159, 104104.

[7] Y. Liu, J. Xiao, J. Koo, B. Yan, Nature materials. **2021**, 20, 638.

[8] S. Alwan, Y. Dubi, J. Am. Chem. Soc. **2021**, 143, 14235.

[9] S. Naskar, V. Mujica, C. Herrmann, J. Phys. Chem. Lett. **2023**, 14, 3, 694.

[10] S. S. Skourtis, D. N. Beratan, R. Naaman, A. Nitzan, D. H. Waldeck, Phys. Rev. Lett. **2008**, 101, 238103.

[11] J. Gersten, K. Kaasbjerg, A. Nitzan, J. Chem. Phys. **2013**, 139, 114111.

[12] M. Kettner, V V. Maslyuk, D. Nürenberg, J. Seibel, R. Gutierrez, G. Cuniberti, K.-H. Ernst, H. Zacharias, J. Phys. Chem. Lett. **2018**, 9, 2025.

[13] T. R. Mcguire, R. I. Potter, IEEE Transaction on Magnetics, **1975**, 11, 1018.

[14] H. Nakayama, M. Althammer, Y.-T. Chen, K. Uchida, Y. Kajiwara, D. Kikuchi, T. Ohtani, S. Geprägs, M. Opel, S. Takahashi, R. Gross, G. E. W. Bauer, S. T. B. Goennenwein, E. Saitoh, Physical Review Lett. **2013**, 110, 206601.

[15] L. K. Zou, Y. Zhang, L. Gu, J. W. Cai, L. Sun, Phys. Rev. B. **2016**, 93, 075309.

[16] T. K. Das, F. Tassinari, R. Naaman, J. Fransson, J. Phys. Chem. C, **2022**, 126, 3257.

[17] M. Kettner, B. Göhler, H. Zacharias, D. Mishra, V. Kiran, R. Naaman, C. Fontanesi, D. H. Waldeck, S. Sęk, J. Pawłowski, J. Juhaniewicz, J. Phys. Chem. C, **2015**, 119, 14542.

[18] J. Fransson, Israel J. Chem. **2022**, 62, e202200046.

[19] J. Fransson, Phys. Rev. B, **2020**, 102, 235416.

[20] L. Zhang, Y. Hao, W. Qin, S. Xie, F. Qu, Phys. Rev. B **2020**, 102, 214303.

[21] M. Barroso, J. Balduque, F. Domınguez-Adame, E. Dıaz, Appl. Phys. Lett **2022**, 121, 143505.

[22] S. Matityahu, Y. Utsumi, A. Aharony, O. Entin-Wohlman, C.A. Balseiro, Phys. Rev. B. **2016**, 93, 075407.

[23] Y. Wu, J. E. Subotnik, Nat. Comm. **2021**, 12, 700.

[24] H.-H. Teh, W. Dou, J. E. Subotnik, Phys. Rev. B. **2022**, 106, 184302.